\documentclass[]{aa}

\usepackage{lineno,xcolor,multirow,array,booktabs,amssymb,amsmath,natbib}%
\usepackage{graphicx,epstopdf,hyperref,appendix,threeparttable,chngpage}
\usepackage{txfonts}
\bibpunct{(}{)}{;}{a}{}{,} 
\begin{document}

   \title{Three-dimensional analyses of an aspherical coronal mass ejection and its driven shock}


   \author{Beili Ying\inst{1,2}\and
          Li Feng\inst{1,2}\and
          Bernd Inhester\inst{3}\and
          Marilena Mierla\inst{4,5}\and
          Weiqun Gan\inst{1,2}\and
          Lei Lu\inst{1}\and
          Shuting Li\inst{1,2}
          }

   \institute{Key Laboratory of Dark Matter and Space Astronomy, Purple Mountain Observatory, Chinese Academy of Sciences, Nanjing 210023, China\\
   \email{lfeng@pmo.ac.cn}
         \and
             School of Astronomy and Space Science, University of Science and Technology of China, Hefei, Anhui 230026, People's Republic of China
         \and
            Max-Planck-Institut f\"{u}r Sonnensystemforschung, G\"{o}ttingen 37077, Lower Saxony, Germany
         \and
            Solar–Terrestrial Centre of Excellence—SIDC, Royal Observatory of Belgium, 1180 Brussels, Belgium
         \and
            Institute of Geodynamics of the Romanian Academy, 020032 Bucharest-37, Romania
             }
    \titlerunning{3D analyses of an aspherical CME and its driven shock}
    \authorrunning{Ying et al.}


  \abstract
   {Observations reveal that shocks can be driven by fast coronal mass ejections (CMEs) and play essential roles in particle accelerations. A critical ratio, $\delta$, derived from a shock standoff distance normalized by the radius of curvature (ROC) of a CME, allows us to estimate shock and ambient coronal parameters. However, true ROCs of CMEs are difficult to measure due to observed projection effects.}
   {We investigate the formation mechanism of a shock driven by an aspherical CME without evident lateral expansion. Through three-dimensional (3D) reconstructions without a priori assumptions of the object morphology, we estimate the two principal ROCs of the CME surface and demonstrate how the difference between the two principal ROCs of the CME affects the estimate of the coronal physical parameters.}
   {The CME was observed by the Sun Earth Connection Coronal and Heliospheric Investigation (SECCHI) instruments and the Large Angle and Spectrometric Coronagraph (LASCO). We used the mask-fitting method to obtain the irregular 3D shape of the CME and reconstructed the shock surface using the bow-shock model. Through smoothings with fifth-order polynomial functions and Monte Carlo simulations, we calculated the ROCs at the CME nose.}
   {We find that (1) the maximal ROC is two to four times the minimal ROC of the CME. A significant difference between the CME ROCs implies that the assumption of one ROC of an aspherical CME could cause overestimations or underestimations of the shock and coronal parameters. (2) The shock nose obeys the bow-shock formation mechanism, considering the constant standoff distance and the similar speed between the shock and CME around the nose. (3) With a more precise $\delta$ calculated via 3D ROCs in space, we derive corona parameters at high latitudes of about -50$^{\circ}$, including the Alfv{\'e}n speed and the coronal magnetic field strength.}
   {}

   \keywords{shock waves -- Sun: corona -- Sun: coronal mass ejections
               }

   \maketitle
%

\section{Introduction}

\bibliographystyle{aa}
Coronal mass ejections (CMEs) are ejections of large magnetized plasma structures. They play essential roles in affecting the solar-terrestrial space and are regarded as primary drivers of geomagnetic storms on Earth. A CME often appears as a typical three-part structure in white-light images with a bright front, a dark cavity, and a dense core \citep{Illing1983}. In observations and theories, the bright classical core is often considered to be a dense filament structure. The bright front can be due to the coronal plasma pileup \citep{Cheng2012, Howard2015b} and the mass supplement by the outflow from dimming regions in the low corona \citep{Tian2012}.

In the field-of-view (FOV) of white-light coronagraphs, a faint front followed by diffuse emissions can often be observed preceding a CME front, a so-called two-front structure \citep{Vourlidas2013, Kwon2014, Feng2020} representing a shock region driven by a fast CME. Shock waves with higher speeds have larger amplitudes \citep{Kienreich2011}. In the solar corona, a fast expansion may act as a three-dimensional (3D) piston; for example, a CME usually expands in all directions to drive a shock \citep{Gopalswamy2011, Kim2012, Ying2018}. On the other hand, one-dimensional (1D) motions may also occur, such as a plasma blob propagating with a constant size and resulting in a shock \citep[e.g.][]{Klein1999}. \citet{Warmuth2007, Warmuth2015} and \citet{Vrsnak2008} have suggested two fundamental processes of shock formation: a bow shock and a 3D piston-driven shock. In the case of a bow shock, when a driver pushes the plasma, the offset distance between the bow shock and the CME is constant in a homogeneous media. Such an offset distance is usually referred to as the ``standoff distance'' and is denoted by $\Delta$; it is affected by the velocity, size, and shape of the driver. In addition, the velocity of the driver is equal to that of the shock. In the context of a 3D piston-driven shock, different characteristics can be distinguished from the bow shock, including an expanding driver pushing the plasma in all directions, the increasing offset distance between the driver and the shock, and distinct velocities (the shock is faster than the piston). This work finds an aspherical CME without evident lateral expansion during the propagation and a CME-driven shock behaving like a bow shock in the nose part. Thus, we investigate the possible shock formation mechanism based on the relationship between kinematic properties of the CME and the shock. Attention is paid to the properties of the CME and the shock nose parts and the relationship between these two structures.

The standoff distance is crucial for inferring the physical parameters of shocks and the coronal magnetic field strength. \citet{Gopalswamy2011} determined the coronal magnetic field strength in the heliocentric distance range 6-23 $R_\odot$ from the measured $\Delta/R_\mathrm{c}$ for a spherical-like CME in the coronagraph FOV. Through the same standoff-distance method, \citet{Kim2012} examined ten round fast limb CMEs to derive coronal physical parameters in the distance range 3-15 $R_{\odot}$ and compared these results with those estimated from the density compression ratio. \citet{Mancuso2019} extended the standoff-distance-ratio (SDR) method from two-dimensional (2D) to 3D using 3D reconstructions of a CME and its driven shock with two spherical surfaces in the inner corona. However, not all CMEs conform to the spherical hypothesis. \citet{Maloney2011} studied the shock driven by a blunt CME using data obtained from the Heliospheric Imager on board the \textit{Solar TErrestrial RElations Observatory} \citep[STEREO;][]{Kaiser2008} and investigated the relationships between the standoff distance and the Mach number through semiempirical equations \citep{Spreiter1966, Farris1994, Russell2002}. They found that their derived SDR versus Mach number cannot match well with the semiempirical equations and noted that the measured radius of curvature (ROC) is underestimated by a factor of $3-8$. They were only able to determine the ROC with considerable uncertainty; therefore, they could not provide a critical test of the theory due to the difficulty in estimating a reasonable $R_\mathrm{c}$, as the observations usually only give a projected value. The CME studied in this work is asymmetric. We attempt to estimate the actual principal ROCs of the 3D CME surface through the 3D reconstruction without the CME topological assumption and demonstrate how the difference between the two principal ROCs of the CME will affect the estimations of the coronal physical parameters. We extend the SDR method from the regular spherical hypothesis of CMEs to an irregular CME. 

This paper presents the multipoint and multiwavelength observations in Sect. 2. Then we show 3D reconstructions of a jet, a CME, and its driven shock in Sect. 3. Section 4 presents the analysis of the morphological and kinematic evolution of the jet, CME, and shock as well as probable shock formation mechanisms. In Sect. 5 we measure the two principal ROCs of the CME and estimate coronal parameters with the SDR method. Finally, discussions and conclusions are given in Sect. 6.

   
 
 \begin{figure*}[!ht]
   \centering
   \includegraphics[width=1.\textwidth]{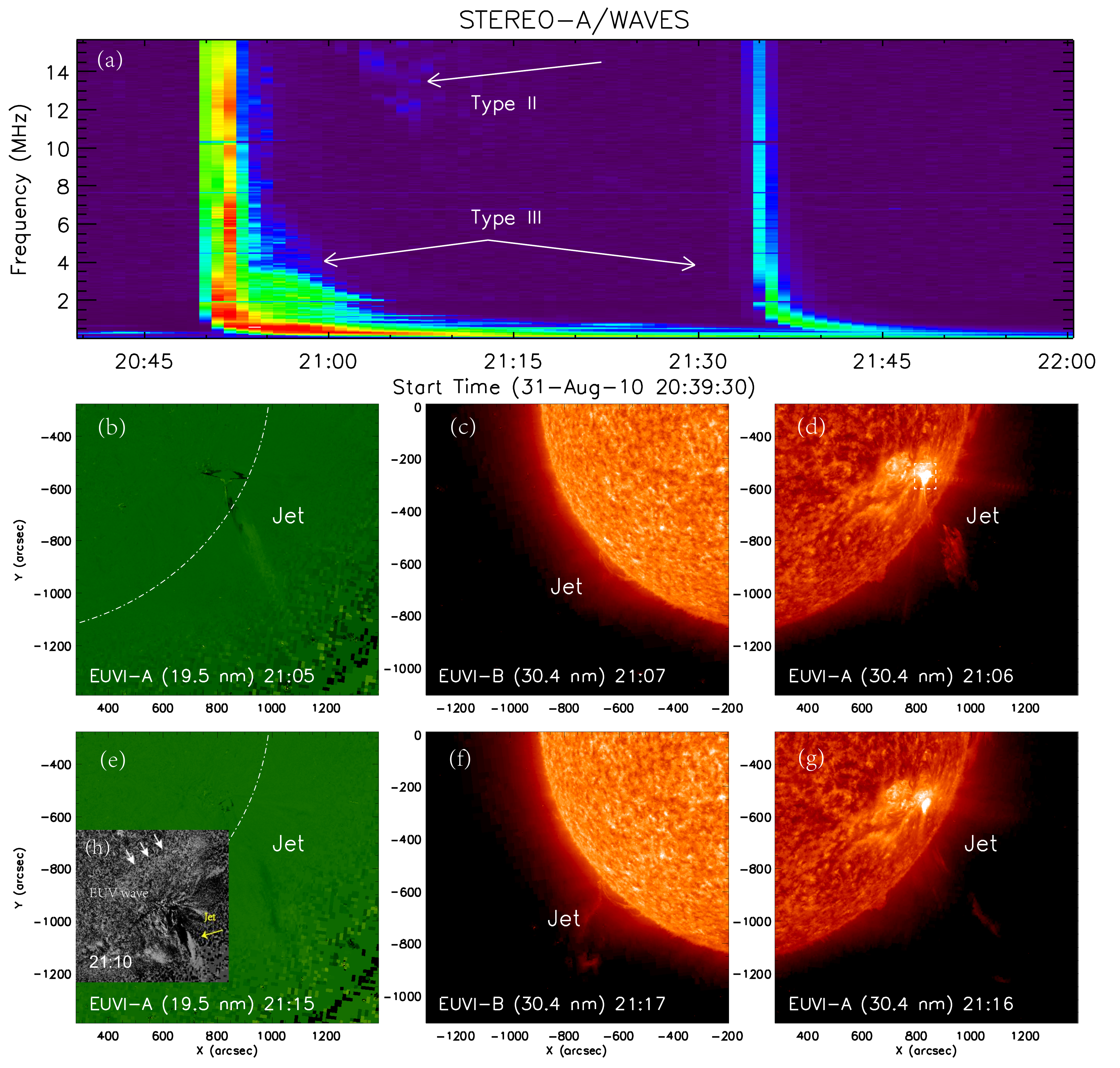}
   
   \caption{Evolution of the jet observed by EUVI on board STA and STB and radio bursts observed by STA/WAVES. (a): STA/WAVES observations of two interplanetary type-III radio bursts as well as a weak and short-lived type-II radio burst, marked by white arrows. (b)-(d): Observations of the jet at $\sim$21:06~UT from STB/EUVI 195 {\AA}, STA/EUVI, and STB/EUVI 304 {\AA}, respectively. The source region is marked by a white box in panel (d). (e)-(g): Jet observations at $\sim$21:16~UT from STB/EUVI 195 {\AA}, STA/EUVI, and STB/EUVI 304 {\AA}. Two running-ratio images (with respect to the previous images) in panels (b) and (e) are chosen to enhance the signal-to-noise of the jet. (h): EUV wave on the disk at $\sim$21:10 UT in STA/EUVI 195 {\AA}.}
   \label{fig:stereo_jet}
 \end{figure*}
 
 \begin{figure*}[!ht]
   \centering
   \includegraphics[width=1.\textwidth]{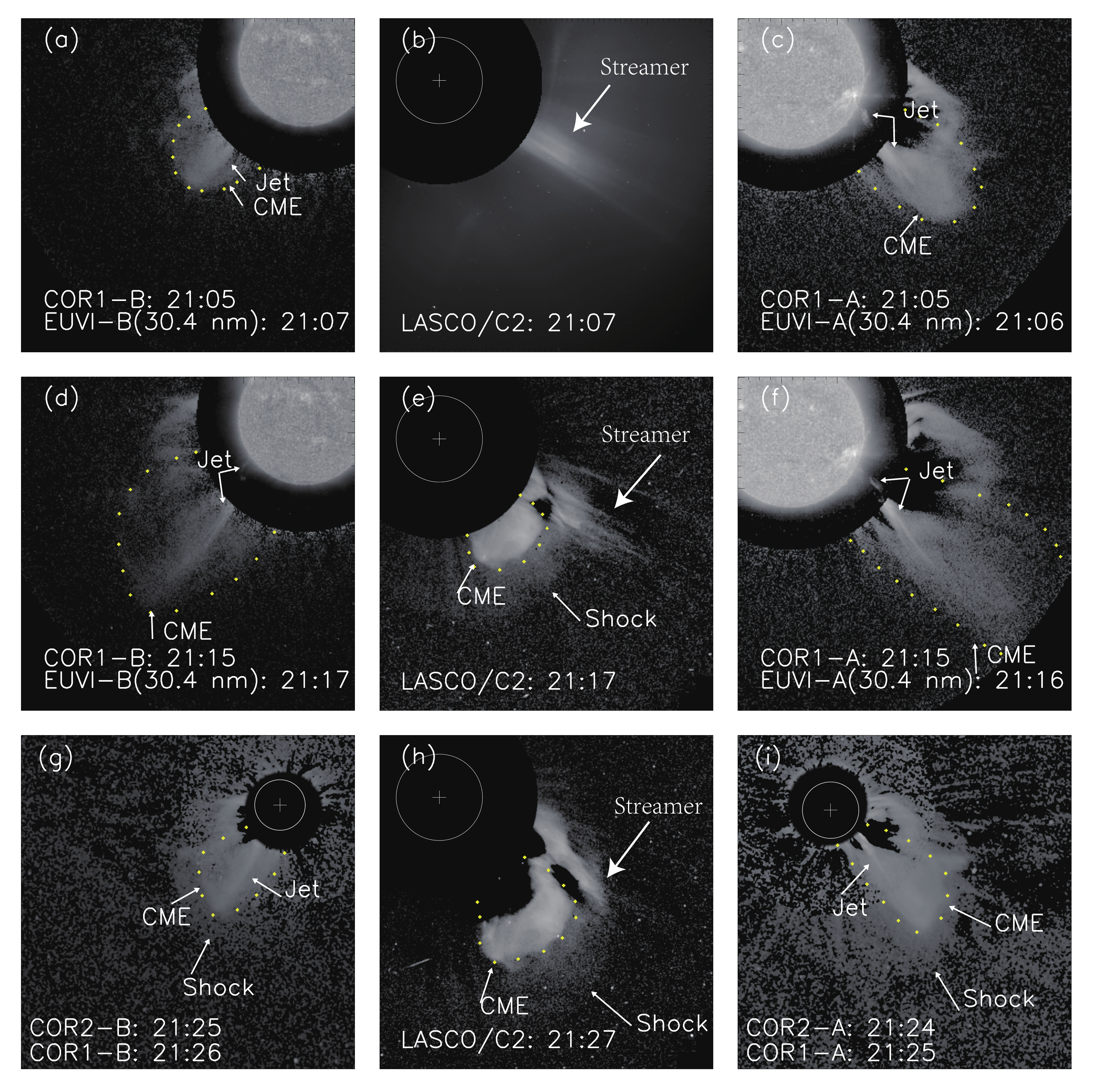}
   \caption{Multipoint observations of the jet, the CME, and its driven shock from the SOHO LASCO/C2 and STEREO EUVI/COR1/COR2 instruments. The three images in each row represent three different perspectives at the proximate time. Each column represents the evolution of events from $\sim$21:05 UT to $\sim$21:25 UT in the same instrument. Panel (b) is an original image. Panels (e) and (h) are running-difference images presented to show the more apparent structure of the shock region. For COR data, the respective pre-CME images are subtracted. The white circle and the plus sign in each panel indicate the limb of the solar disk and solar center. The small yellow plus signs mark the possible CME peripheries. Arrows denote the jet, CME, and shock.}
   \label{fig:multi_point_obs}
 \end{figure*}
 
 
 \begin{figure*}[!th]
   \centering
   \includegraphics[width=1.\textwidth]{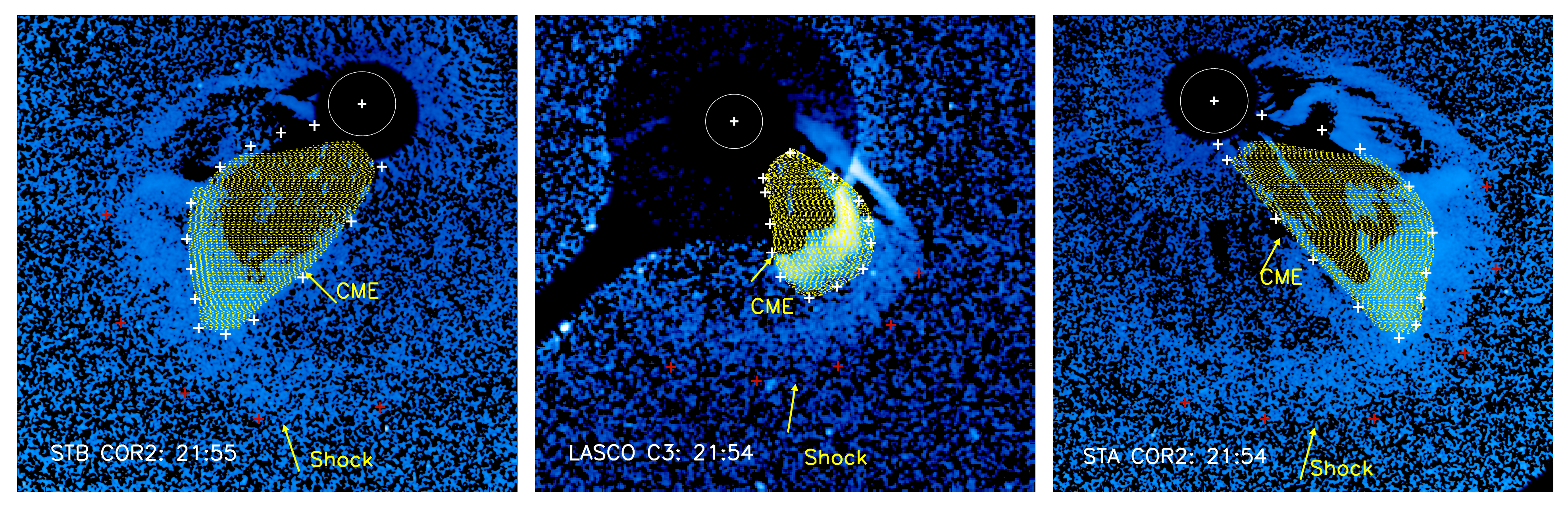}
   \includegraphics[width=1.\textwidth]{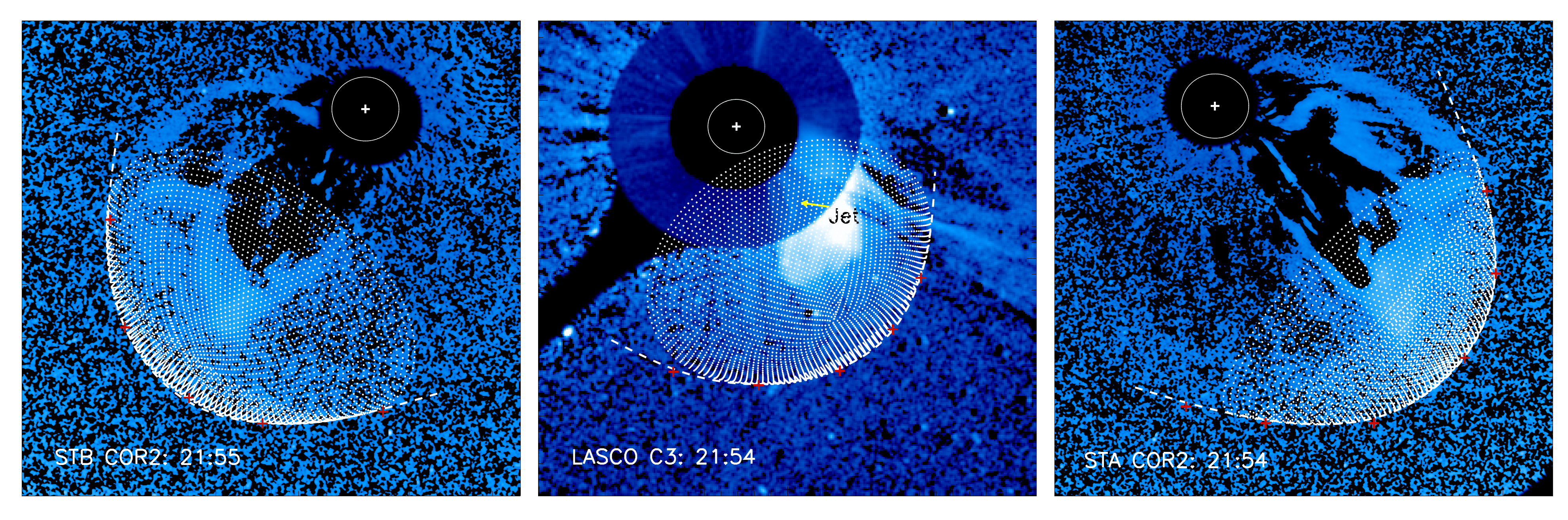}
   \caption{Projections of the 3D CME and shock in the three perspectives. Top: Multipoint running-difference images at $\sim$21:54 UT. Yellow grids in the three perspectives mark projections of the 3D CME obtained from the MF method. The small white and red plus signs mark the possible peripheries of the CME and the shock, respectively. Bottom: Projections of the 3D shock surface reconstructed by the bow-shock model on LASCO/C3 and STEREO/COR2 images. The bottom-middle panel is the base-difference image. The dashed white lines represent the position of the extended bow-shock surface in three perspectives. The red plus signs have the same positions as those in the top panels. Arrows mark the CME, shock, and jet. }
   \label{fig:reconstrucutre}
 \end{figure*}

\section{Observations}


An eruptive event occurred on 31 August 2010, which was accompanied by jets and two concurrent CMEs from an active region (W67$^{\circ}$, S23$^{\circ}$) located close to the west solar limb in the FOV of STEREO-A (STA). A two-front structure was observed in the FOV of the white-light coronagraphs COR1 (FOV: 1.4-4 $R_{\odot}$) and COR2 (FOV: 2.5-15 $R_{\odot}$) in the Sun Earth Connection Coronal and Heliospheric Investigation \citep[SECCHI;][]{HowardRA2008} instrument suite on board STEREO and the Large Angle Spectroscopic Coronagraph \citep[LASCO;][]{Brueckner1995} C2 (FOV: 2-6 $R_{\odot}$) and C3 (FOV: 3.7-30 $R_{\odot}$) on board \textit{Solar and Heliospheric Observatory} \citep[SOHO;][]{Domingo1995}.

A jet that first erupted at around 20:50 UT was captured by the STA/WAVES instruments; it was determined to be a type-III radio burst with a fast frequency-drift rate (Fig.~\ref{fig:stereo_jet} (a)). Figures~\ref{fig:stereo_jet} (b)-(g) show the evolution of the jet from $\sim$21:05 UT to $\sim$21:15 UT from the two perspectives of STA and STEREO-B (STB). The eruptive source region in the FOV of the extreme ultraviolet imager on board STA (EUVI-A) is shown in Fig.~\ref{fig:stereo_jet} (d) in 304 {\AA} passband. Two running-ratio images (with respect to the previous images) chosen to display clearer structures of the jet in the EUVI-A 195 {\AA} passband are shown in Figs.~\ref{fig:stereo_jet} (b) and (e). The jet had already appeared in the FOV of STEREO COR1 at around 21:05 UT (marked by white arrows), as shown in Fig.~\ref{fig:multi_point_obs}.  

In Fig.~\ref{fig:stereo_jet} (a), a short-lived type-II radio burst in the dynamic spectra of WAVES on board STA indicated that a shock formed at around 21:02 UT. Then, we started to see wave structures on the disk at $\sim$21:05 UT and most clearly at 21:10 UT (Fig.~\ref{fig:stereo_jet} (h)) in the EUVI-A 195 {\AA} images. Recently, \citet{Kouloumvakos2021} demonstrated that a strong, supercritical, quasi-perpendicular shock wave could lead to type-II emission through 3D shock modelling and radio observations. They also found that the shock can be formed before the start of the type-II radio burst and that the type-II radio burst will end when the geometry of the shock is oblique to quasi-parallel, even if the shock has a strong compression ratio. The patch shape of the type-II burst might be due to the sensitive dependence of the radio flux on the shock and solar wind properties  \citep{Knock2003}. The CME with a projected velocity of 1300 km s$^{-1}$, provided by the Coordinated Data Analysis Workshop, came into the FOV of LASCO/C2 at around 21:17 UT, as shown in Fig.~\ref{fig:multi_point_obs} (e). Its actual speed is measured using 3D reconstructions in Sect. 4. The high speed of the CME indicates that a shock can be driven, which is further evidenced by the more diffuse front in the coronagraph images shown in Figs.~\ref{fig:multi_point_obs} (g)-(i). Many studies have revealed that the diffuse emission on the periphery of the brighter CME front can represent the shock signature \citep{Vourlidas2003,Vourlidas2013,Ontiveros2009}. It should be noted that two CMEs are seen in the LASCO images. The CME that we are interested in is delineated by plus signs in Fig.~\ref{fig:multi_point_obs}. The properties of the second CME, to the north of our CME and seen below the streamer, are beyond the scope of this paper. Figures~\ref{fig:multi_point_obs} (e) and (h) are running-difference images created to enhance the signal-to-noise ratio of the shock region, while the remaining images are base-difference images subtracted from the pre-event images at $\sim$20:35 UT for COR1 and at $\sim$20:40 UT for COR2 observations, except for the original LASCO/C2 image in Fig.~\ref{fig:multi_point_obs} (b). In Fig.~\ref{fig:multi_point_obs}, all structures (i.e., the jet, CME, and shock region) are marked. In the FOV of the LASCO/C3, we can see a small deflection of the CME and a severe bend of the jet with a ``J'' shape, as shown in Fig.~\ref{fig:reconstrucutre}.  One possible explanation for the CME deflection is that it may interact with coronal holes and/or solar-wind stream boundaries when propagating outward and can be affected by ambient flows of different origins \citep{Luhmann2020SoPh}. 

   
 
 \begin{figure*}[!th]
   \centering
   \includegraphics[width=1.\textwidth]{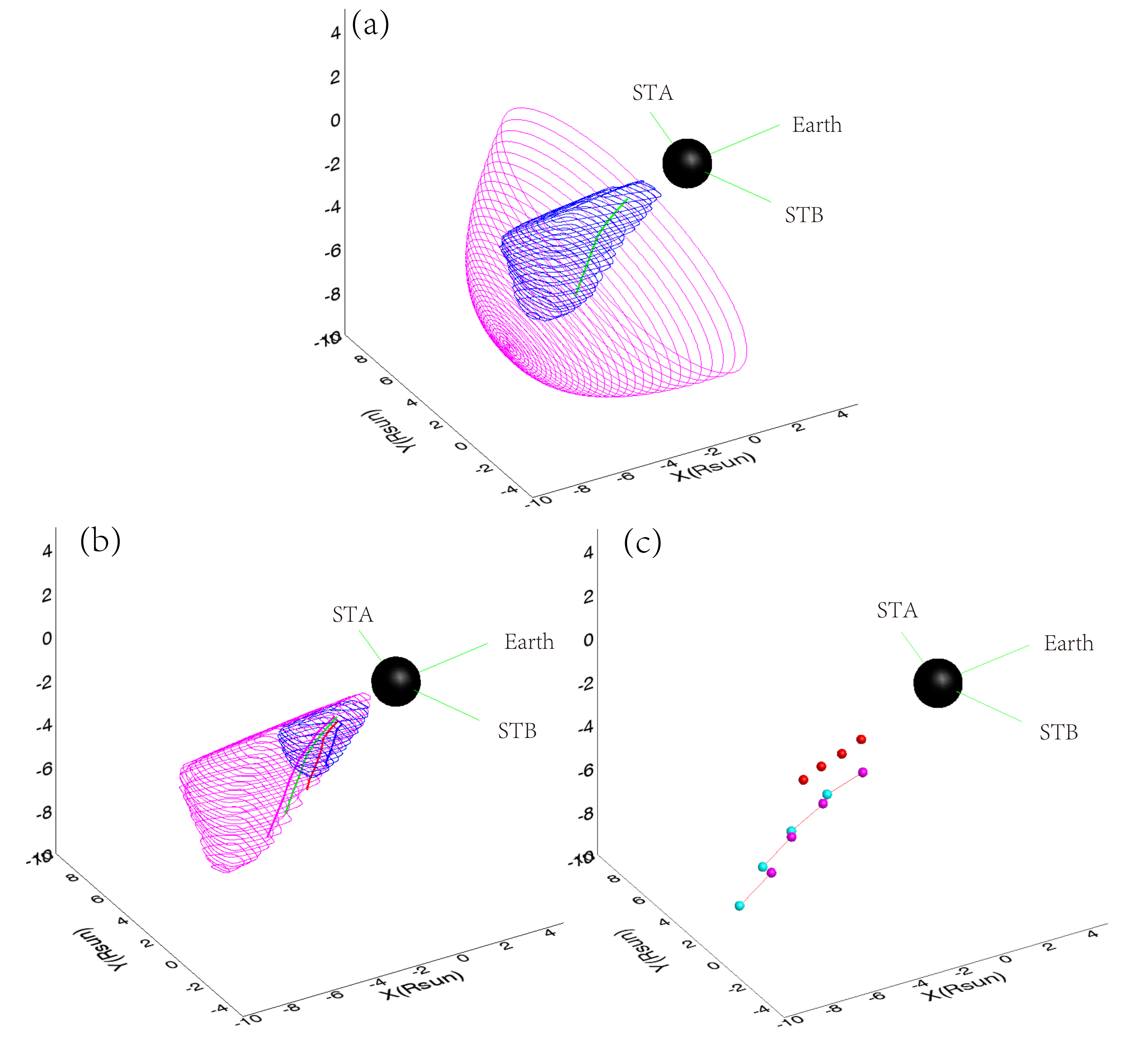}
   \caption{3D reconstructions of the CME, jet, and shock. (a): One example of a 3D jet (green), a CME cloud (blue), and a shock surface (magenta) at 21:55 UT. The three thin green axes show the orientations of STA, Earth, and STB, in a clockwise direction. (b): CME observed at 21:25 and 22:10 UT (the two clouds) and the jet observed at 21:25, 21:40, 21:55, and 22:10 UT (the four thick solid lines of different colors). (c): GCs of the CME clouds (red dots) and the CME nose at 21:25, 21:40, 21:55, and 22:10 UT (magenta dots). Cyan dots indicate the shock apex, and the orange line denotes the symmetrical bow axis pointing from the CME nose to the shock apex.}
   \label{fig:3d_jet_shock}
   
 \end{figure*}
 \section{3D reconstructions of the jet, CME, and shock}

 Multi-perspective observations allow us to reconstruct the jet, CME, and shock and analyze their natural characteristics in 3D space, avoiding the effect of projections. On 31 August 2010, STA and STB had separation angles with Earth, of 81 and 74 degrees, respectively. Concerning the 3D geometrical reconstruction of the CME location, different techniques have been developed: (1) forward modelings \citep{Thernisien2006, Wood2009}, (2) tie-pointing plus triangulation methods \citep{deKoning2009, Howard2008, Liewer2009,Li2018JGRA,Li2020JGRA,Lyu2020AdSpR}, and (3) polarization ratio methods \citep{Mierla2009SoPh,Moran2010, Lu2017}. Details of these methods can be found in the review by \citet{Mierla2010}.

 In this work, we reconstructed the 3D geometry of the CME by utilizing a method proposed by \citet{Feng2012b, Feng2013a}, the so-called mask fitting (MF) method. The MF technique involves finding the best 3D CME shape that fits the traced CME peripheries from different viewpoints using projections and smoothings. It allows us to obtain the 3D shape of a CME cloud without assuming a predefined family of shape functions and to reveal more details, such as the different ROCs in different planes.  One example of a 3D CME cloud, at 21:55 UT, is shown in Fig.~\ref{fig:3d_jet_shock} (a). We defined the farthest point of the CME bulge from the Sun as the CME nose. Projections of the CME into the image planes from three perspectives are shown in the top panels of Fig.~\ref{fig:reconstrucutre}. Due to the occulter of LASCO, a small part of the reconstructed CME near the solar equator is missing. The missing part will not affect our analysis when attention is paid to the CME nose part.

 Given the elongated shape of the jet, we were able to reconstruct its 3D morphology by using scc\_measure.pro, which is available via Solar Software. As the jet is more pronounced in the LASCO coronagraph images, we started marking the position of the jet in the LASCO C2 and C3 images. Scc\_measure.pro produced the epipolar lines \citep{Inhester2006} in the STA COR1 and COR2 images to guide the selection of the corresponding pixels. Once the correspondence between pixels was found, the 3D reconstruction was achieved by calculating lines of sight that belong to the respective pixels in images and back-projecting them into the 3D space. This procedure returns the 3D coordinates of the selected jet pixels. We also projected the 3D points into the plane of STB at the same time to check whether these 3D reconstructed points are located in the jet structure. One example of a 3D jet is shown in Fig.~\ref{fig:3d_jet_shock} (a).

 Considering the much smoother bow-shock shape around the shock nose, we reconstructed the shock surface under the assumption of a symmetrical 3D bow-shock geometry \citep[][]{Ontiveros2009, Chenyao2014}. The bow-shock surface is given by \citep{Smith2003}
 \begin{equation}
   Z(X,Y)=h-\frac{d}{s}\times(\frac{\sqrt{X^2+Y^2}}{d})^s ,  
   \label{eq:bowshock} 
 \end{equation}
 where $h$ is the shock apex height from the CME nose (marked by point ``$O$''), $s$ controls the bluntness of the shock, and $d$ controls the width of the shock shape. The Z axis points from the CME nose to the shock apex, and $X$ and $Y$ are the coordinates in the plane perpendicular to the Z axis. The equation is given in a local Cartesian coordinate system \citep[more details are available in the appendix of ][]{Chenyao2014}. An example of a 3D shock is shown in Fig.~\ref{fig:3d_jet_shock} (a). The projections of the reconstructed shock surface into the image planes from three perspectives are shown in the bottom panels of Fig.~\ref{fig:reconstrucutre}.


\begin{figure}[!th]
   \centering
   \includegraphics[width=0.48\textwidth]{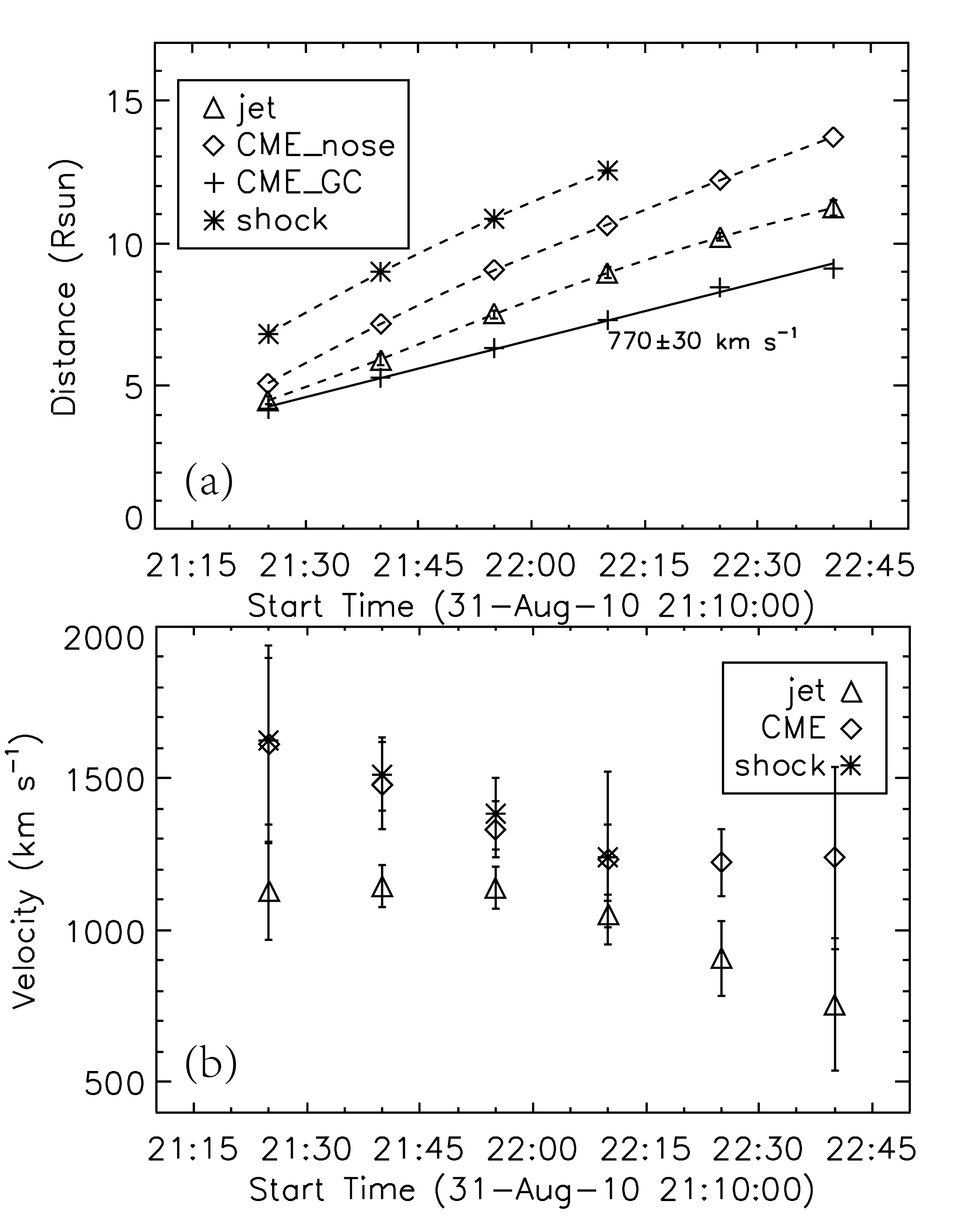}
   \caption{Kinematics evolution of the jet, the CME, and the shock. (a): Time-distance plot of the jet (triangles), CME GC (plus signs), CME nose (diamonds), and its driven shock (asterisks).  The corresponding dashed lines are the smoothing cubic spline fittings to the time-distance plots. The solid line denotes the linear fit of the CME GC distance. (b): 3D speeds of the jet, CME nose, and shock. }
   \label{fig:3d_dis_speed}
 \end{figure}
 \section{Morphological and kinematic evolution of the jet, CME, and shock} 
 This section presents the 3D morphological evolution of the jet, the CME, and the shock. Subsequently, we analyze the kinematic properties of the CME nose and shock to explore the relationship between the CME and shock and the formation mechanism of the shock around the nose.

 \subsection{Morphological evolution}
 The evolution of the 3D CME, with the CME cloud observed at 21:25 UT and 22:10 UT, is shown in Fig.~\ref{fig:3d_jet_shock} (b). The CME shows a nearly constant angular width without evident lateral expansion during its evolution. However, noticeable morphological changes occur in the southern part of the CME front, which varies from a smooth surface to a bulge. The evolution of the jet inside the CME at four time instances is shown in Fig.~\ref{fig:3d_jet_shock} (b). The jet's translational motion, most prominently seen from the Earth view, might be due to the slingshot effect of the strongly curved magnetic field caused by magnetic reconnection \citep[e.g.][]{ShenYD2011}. 
 
 Figure~\ref{fig:3d_jet_shock} (c) displays the difference between overall and local trends of the CME and the relationship between the CME and shock noses. The CME cloud's geometric centers (GCs; see \citealt{Feng2012b} for definitions of GCs) are shown at 21:25, 21:40, 21:55, and 22:10 UT. We note that the CME GC only takes account of what is peaking beyond the occulter boundary. Thus, the position of the GCs, which are used to roughly indicate the overall trend of the CME, will be higher than their true heights. The main propagation direction of the CME is kept almost constant. In contrast, the CME nose obviously bends. Consequently, the shock apex and its symmetrical axis also change apparent direction. The average latitude of the CME nose and the shock apex is about -50 degrees.
\subsection{Kinematic evolution}

In Fig.~\ref{fig:3d_dis_speed} (a) the distance evolution of the jet, CME GC, CME nose, and shock nose is presented. The jet positions are measured from scc\_measure.pro. Uncertainties in deriving the jet distance were estimated by manually marking the jet front ten times at each time instance. Such uncertainties were then propagated to the speed estimate of the jet. The standoff distance between the CME nose and shock apex keeps an almost constant value of 1.8 R$_{\odot}$, as shown in \autoref{tab:01} (Col. 7). The CME GC propagates outward with an average speed of 770 $\pm$ 30 km s$^{-1}$, which is much slower than the speed of the CME nose, as shown in Fig.~\ref{fig:3d_dis_speed} (b). Within the error bars, the speed of the CME nose is consistent with the shock speed from 21:25 UT to 22:10 UT in Fig.~\ref{fig:3d_dis_speed} (b). The uncertainty of these speeds mainly comes from the distance measurement from the 3D reconstruction. We used 100 Monte Carlo simulations to obtain the speed errors. In such simulations, 100 values following a Gaussian distribution were chosen randomly with 3$\sigma =$0.72 R$_{\odot}$ for the shock and CME distance. The derived error bars are plotted in Fig.~\ref{fig:3d_dis_speed} (b). From 21:25~UT to 22:10~UT, when the CME and shock noses are visible in the coronagraph images from three perspectives, both the shock and CME speeds drop from $\sim$1600 km s$^{-1}$ to $\sim$1200 km s$^{-1}$. 
\subsection{Shock formation mechanism}

Considering the constant standoff distance ($\Delta\sim1.8~R_{\odot}$), the similar speeds of the CME nose and the shock apex, and the bow-shock shape of the shock nose, we infer that the nose part of the shock forms predominantly due to the fast propagation of the driving CME and resembles a bow-shock mechanism. We note that only in the homogeneous ambient medium would the driven bow shock have precisely the same speed as its driver. The anisotropic distribution of the solar wind may affect the properties of the bow shock \citep{Vrsnak2008}. As shown in the right panels of Fig.~\ref{fig:reconstrucutre}, with respect to the symmetric axis of the shock, the eastern flank of the shock generally follows the bow-shock shape. Moreover, the corresponding eastern flank of the CME does not show evident expansion, which rules out the piston-driven mechanism due to expansion. The shock on the western flank, close to the equator, is behind the reconstructed bow shock. The lack of a bow-shock wave on the western flank might be because the wave passed through a high-density streamer and a slow solar wind regime at low latitude. As shown in Fig.~\ref{fig:multi_point_obs} (b), there is a high-density streamer before the eruption, which can be seen clearly in the original STA COR1 and COR2 images as well. According to the adiabatic shock equation \citep[][Chapter VII]{Gurnett2017}, a lower Alfv\'{e}n speed (such as in the streamer) and slower background solar wind usually result in a lower shock propagation speed. 

\begin{table*}[htb]

   \centering
   \caption{Properties of the shock, CME, and the ambient medium when assuming $\gamma = 4/3.$}
   \label{tab:01}
   \begin{threeparttable}
   \resizebox{18. cm}{21 mm}{ 
   \raggedright
   \begin{tabular}{ccccccccccc}
   \toprule
  
   \multirow{3}{*}{Time} & \multicolumn{4}{c}{CME parameter} & \multicolumn{3}{c}{Shock parameter} & \multirow{3}{*}{M} & \multirow{3}{*}{$V_{A}$} & \multirow{3}{*}{B} \\ \cmidrule(r){2-5} \cmidrule(r){6-8}
 & $D_{nose}$ & {$ R_{\rm min}$} & {$ \bar{R}_c$} & {$ R_{\rm max}$} & $D_{sh}$ & $\Delta$ & $\Delta/R_c$ &  &  &  \\
(UT) & ($\rm R_{\odot}$) & ($\rm R_{\odot}$) & ($\rm R_{\odot}$) & ($\rm R_{\odot}$) & ($\rm R_{\odot}$) & ($\rm R_{\odot}$) &  &  & ($\rm km~s^{-1}$) & (mG) \\ 
   \midrule
   
   21:25 &5.10 & 0.37$\pm$0.02 & 0.47$\pm$0.03  & 0.69$\pm$0.07 & 7.10 & 1.80 &  3.79$\pm$0.22 & 1.10$\pm$0.02 & 890$\pm$13 & 43.80$\pm$0.23 \\
   21:40 &7.18 & 0.28$\pm$0.03 & 0.42$\pm$0.05  & 0.84$\pm$0.23 & 9.27 & 1.80 &  4.33$\pm$0.45 & 1.10$\pm$0.01 &  780$\pm$20 & 27.18$\pm$0.26 \\
   21:55 &9.07 & 0.27$\pm$0.01  & 0.42$\pm$0.02  & 1.01$\pm$0.14 &  11.16 & 1.80 &  4.29$\pm$0.22 & 1.10$\pm$0.01 &  660$\pm$10 & 18.38$\pm$0.22 \\
   22:10 &10.62 & 0.43$\pm$0.05  & 0.57$\pm$0.05 & 0.87$\pm$0.13 &  12.81 & 1.90 &  3.35$\pm$0.29 &1.11$\pm$0.03 &  520$\pm$13 & 12.47$\pm$0.14 \\
   22:25 &12.21& 0.28$\pm$0.01 & 0.42$\pm$0.02 & 0.88$\pm$0.09   &   $-$  & $-$  & $-$ &  $-$  &  $-$  &  $-$   \\
   22:40 &13.71& 0.51$\pm$0.05  & 0.68$\pm$0.06  & 1.05$\pm$0.23 &   $-$  & $-$  & $-$ &  $-$  &  $-$  &  $-$   \\
   \bottomrule
   \end{tabular}}
   
   
      \tablefoot{{Col. 1: Observation time. Col. 2: The heliocentric distance of the CME nose. Cols. 3-5: $R_{\rm min}$, $\bar{R}_c$, and $ R_{\rm max}$ are the minimal, mean, and maximal ROCs of the CME nose, respectively. Uncertainties of ROCs are computed~with 100 Monte Carlo simulations. Col. 6: The heliocentric distance of the shock apex. Col. 7: The standoff distance ($\Delta$) between the CME nose and the shock apex from the bow-shock fitting. Col. 8: The SDR, $\delta$, normalized by $\bar{R}_c$. Cols. 9-11: The Alfv{\'e}n Mach number, Alfv{\'e}n speed, and magnetic field strength of the ambient corona calculated with $\bar{R}_c$.}
      }
   \end{threeparttable}
\end{table*}

\section{CME radii of curvature and coronal parameters}

Many previous studies have revealed that geometrical relations between CMEs and shocks can be used to infer coronal parameters \citep{Gopalswamy2011,Kim2012,Mancuso2019} according to a formula based on a model built by \citet{Farris1994} and further extended by \citet{Russell2002}:
\begin{equation}
  \delta=\frac{\Delta}{R_c}=0.81\frac{(\gamma-1)M^2+2}{(\gamma+1)(M^2-1)},
  \label{eq:Mach}
\end{equation}
where $\Delta$ is the standoff distance between the CME front and its driven shock, $R_c$ is the obstacle ROC at the CME nose, $\gamma$ is the ratio of specific heats, and $M$ is the Alfv{\'e}n Mach number. The Alfv{\'e}n speed is more considerable than the local sonic speed in the corona.  

In this work, the standoff distance, $\Delta$, was obtained from 3D reconstructions (in \autoref{tab:01}, Col. 7). If the ROC of the CME can be determined, it is easy to derive the Alfv{\'e}n Mach number based on Eq.~\ref{eq:Mach}. 

\subsection{Two principal radii of curvature of the CME}

Due to the limitations of the observations, many authors have only been able to measure the ROC of CMEs in the plane-of-sky \citep[POS;][]{Gopalswamy2011, Gopalswamy2012, Kim2012} by assuming that a CME possesses an axial-symmetric (e.g., spheric) shape \citep{Mancuso2019}. \citet{Maloney2011} found that the derived SDR versus Mach number cannot match well with the theoretical results derived by Eq.~\ref{eq:Mach} and that the measured ROC is underestimated by a factor of $3-8$. Some CMEs may have an asymmetric shape, and, at a given point on their surface, there are two principal (maximal and minimal) ROCs with different values. These two ROCs could couple with each other and result in a complex effect on the shock's standoff distance, $\Delta$. The mean curvature is defined as the average of the two principal curvatures, and the mean ROC is the reciprocal of the mean curvature. If a CME has an irregular morphology, using only one ROC in the POS may lead to either an overestimation or underestimation of the Alfv{\'e}n Mach number.

In this work, to calculate the ROCs at the CME nose that we are particularly interested in, we combined smoothings with fifth-order polynomial functions and Monte Carlo simulations. The former is to smooth the reconstructed surface of the CME bulge, and the latter is to estimate the uncertainties of ROCs propagating from the surface fitting. Details of the surface fittings for smoothing and the calculations of minimal, mean, and maximal ROCs are described in Appendices A and B, respectively. 

The measured minimal, mean, and maximal ROCs are listed in \autoref{tab:01} (Cols. 3-5). In our event, the averaged value of $\bar{R}_c$ is $\sim$0.5 $\rm R_{\odot}$ at the CME nose (Cols. 4), and this value does not change significantly during the CME propagation. Furthermore, the maximal ROC is about two to four times the minimal ROC. The ratio of the CME ROCs estimated in this article is similar to that measured by \citet{Maloney2011}, which highlights that simply using the CME ROC in the POS could give rise to overestimations or underestimations of the coronal parameters when using the SDR method, especially for events with irregular morphologies. Thus, multi-perspective observations and 3D reconstructions are necessary to obtain more accurate measurements of the properties of solar eruptions (e.g., CMEs and shocks) and coronal physical parameters.

\subsection{Coronal parameters}
\begin{figure}[!th]
   \centering
    \includegraphics[width=0.48\textwidth]{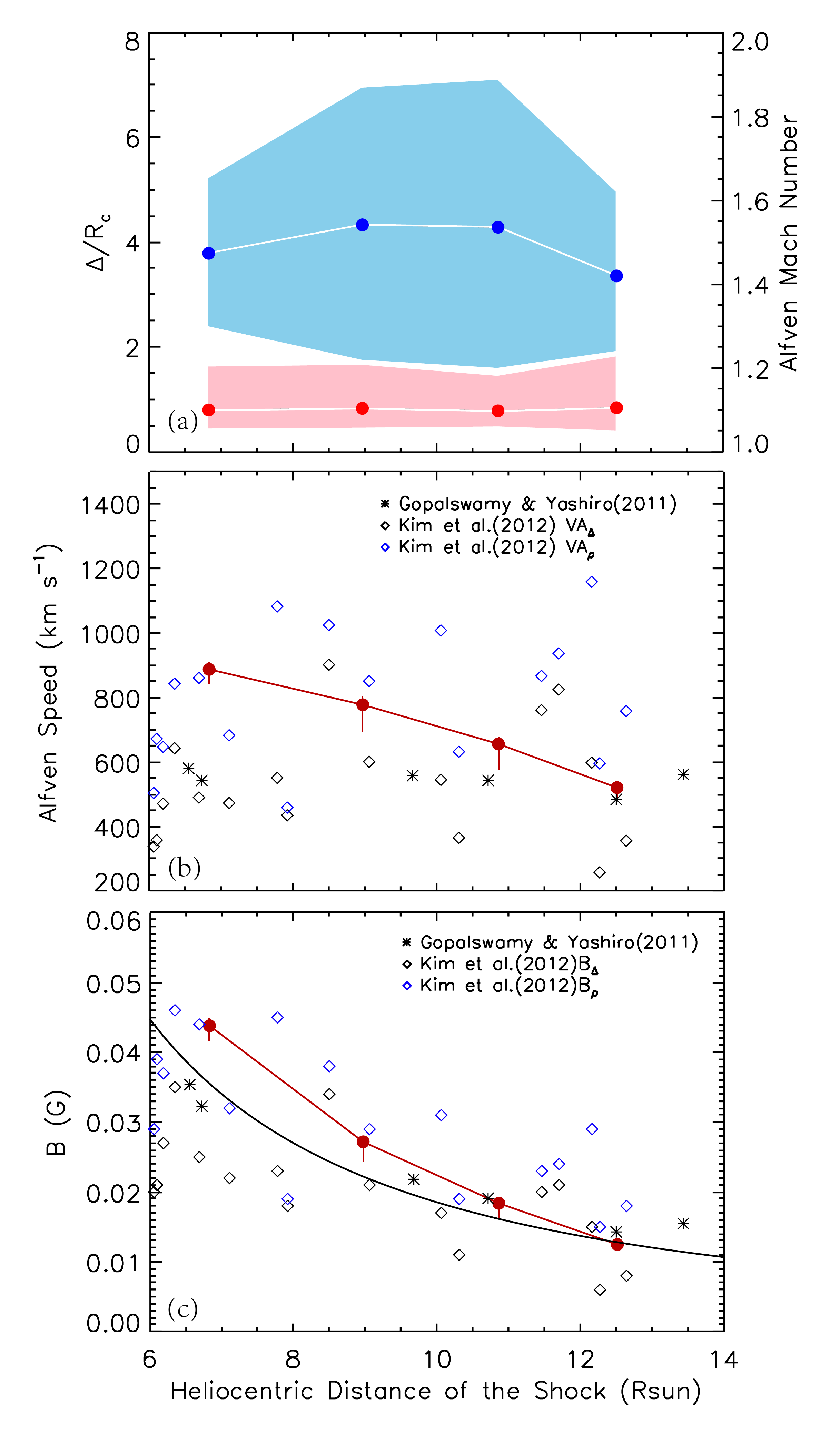}
   \caption{Property evolution of the shock and the ambient medium. (a): Ratios $\delta$ (blue) of the standoff distances ($\Delta$) normalized by ROCs and Alfv{\'e}n Mach numbers, $M$ (pink). The shaded areas denote $\delta$ normalized by the minimal and maximal ROCs plus the errors propagated from surface fittings and the errors of $M$ propagated from $\delta$. (b): Alfv{\'e}n speeds, $V_{A}$, for $\gamma =$4/3. The errors of $V_{A}$ are propagated from $M$ and the shock velocities. The $V_A$ derived by \citet{Gopalswamy2011} (asterisk) and \citet{Kim2012} (diamond) are marked. (c): Coronal magnetic field strength, $B$, for $\gamma =$4/3, calculated using the SDR method. The filled circles represent $B$ derived from the density model of \citet{Leblanc1998}. The solid line is the empirical magnetic field model in \citet{Dulk1978}. The $B$ estimated in \citet{Kim2012} and \citet{Gopalswamy2011} are marked by diamonds and asterisks.}
   \label{fig:standoff}
 \end{figure}

Figure~\ref{fig:standoff} (a) shows the evolution of the SDR, $\delta=\Delta/R_c$ (normalized by the different ROCs at the CME nose), with heliocentric distance. The upper and lower limits of the blue shaded area are $\delta$ calculated with the standoff distance normalized by the minimal and maximal ROCs plus the measured uncertainties propagated from the surface fittings. The Alfv{\'e}n Mach number is derived from Eq.~\ref{eq:Mach} under the assumption of the ratio of specific heat $\gamma$= 4/3. The average of the Alfv{\'e}n Mach number derived from the mean ROC is $\sim$1.1. The low Alfv{\'e}n Mach number might be the reason for the type-II radio-quiet burst \citep{Kouloumvakos2021}. The Alfv{\'e}n speed is $V_A=\frac{V_{sh}-V_{sw}}{M}$, where $V_{sh}$ is the shock speed shown in Fig.~\ref{fig:3d_dis_speed} (b) and $V_{sw}$ indicates the solar wind velocity. In this work, we selected the solar wind speed produced by a 3D magnetohydrodynamics solar wind model at the solar minimum year \citep[Fig. 2 of][]{Jin2012,vanderHolst2010} to denote the possible solar wind speed for this event. The corresponding solar wind speeds are 500, 575, 625, and 650 km s$^{-1}$ from 21:25 UT to 22:10 UT at a high latitude of around -50 degrees. As shown in Fig.~\ref{fig:standoff} (b), the Alfv{\'e}n speed decreases from $\sim$890 km s$^{-1}$ to $\sim$520 km s$^{-1}$. Based on the magnetic field extrapolated from data from the Helioseismic and Magnetic Imager  on board the \textit{Solar Dynamics Observatory} (SDO) and the electron number density derived from SDO/AIA (the Atmospheric Imaging Assembly) and SOHO/LASCO, \citet{Zucca2014} computed the Alfv{\'e}n speed distribution. They found that $V_A$ is probably larger at higher latitudes than around the equator. This is consistent with our findings. The obtained Alfv{\'e}n speeds also fall in the range of the Alfv{\'e}n speed in \citet{Kim2012} derived with the SDR method and the density-compression-ratio method in Fig.~\ref{fig:standoff} (b), respectively. 

The upstream magnetic field strength, $B$, can be determined by using
 \begin{equation}
  B=5\times 10^{-5} V_A n^{1/2},
  \label{eq:magnetic}
\end{equation}
where $n$ is the upstream plasma number density in units of $\rm{cm^{-3}}$. We chose a density model from \citet{Leblanc1998} to derive $n$. The formula is 
\begin{equation}
  n_{[10^8 \rm cm^{-3}]}=\frac{0.8}{r^6}+\frac{0.041}{r^4}+\frac{0.0033}{r^2},
  \label{eq:model2}             
\end{equation}
where $r$ is the heliocentric distance. The corresponding magnetic field strengths and the empirical relation for $B$ above active regions derived in \citet{Dulk1978} are shown in Fig.~\ref{fig:standoff} (c).  For comparisons, we also include the magnetic field strengths calculated by \cite{Gopalswamy2011} and \citet{Kim2012}.

\section{Discussions and conclusions}
An unusual aspherical CME without evident lateral expansion, which drives a bow shock at the nose, occurred on 31 August 2010. Based on multi-perspective observations, including from STA, STB, and SOHO, we have performed 3D reconstructions of the jet, CME, and its driven shock using different methods. The elongated jet was reconstructed using tie-pointing and triangulation methods, the irregular 3D CME shape was obtained via the MF method without a priori assumptions regarding the object morphology, and the 3D shock was reconstructed via the bow-shock model \citep{Smith2003}. The reconstructed results allow us to analyze these structures' kinematic and morphological evolution in space and investigate the shock formation mechanism. We measured the two principal ROCs of the CME nose via 3D reconstructions for the first time and extended the SDR method from the spheric hypothesis of CMEs to an irregular 3D CME. The main conclusions are summarized below:

(1) CME ROCs: The maximal ROC of the CME nose is about two to four times the minimal ROC, with the mean ROC around 0.5\,$\rm R_{\odot}$. The estimated factor is similar to the factor of $3-8$ estimated by \citet{Maloney2011}. This is the first time such detailed 3D analyses of the CME shape have been carried out to measure the actual ROCs of an aspherical CME. Therefore, such ROCs are crucial for deriving other relevant physical quantities, and 3D reconstructions can improve the measurements of coronal physical parameters.

(2) CME kinematics: The CME's main direction and angular width remain constant, indicating that the CME flank does not expand during the propagation, while the CME nose bends further, causing the deflection of the shock apex and its symmetrical axis. The speed of the CME nose (the farthest CME point from the Sun) reaches up to $\sim$1600\,$\rm km\,s^{-1}$ and drops to $\sim$1200\,$\rm km\,s^{-1}$ at the distance from 5 to 14\,$\rm R_{\odot}$.

(3) Shock formation: Due to the high speed of the CME nose, a shock formed after 21:02 UT. Given the bow-shock shape of the shock nose, we used the bow-shock model to reconstruct the 3D shock and obtain the standoff distance, $\Delta$, between the shock apex and the CME nose. We find that $\Delta$ keeps a nearly constant value of around 1.8 $\rm R_{\odot}$. Based on the almost constant $\Delta$, the similar speeds of the shock and CME noses, the similar mean ROC of the CME with time, and the bow-shock shape of the shock nose, we infer that the nose part of the shock is consistent with the bow-shock scenario described by \citet{Vrsnak2008} and \citet{Warmuth2007, Warmuth2015}. The reconstructed bow shock can also generally fit the shock's eastern flank observed by STA with respect to the symmetric shock axis. In contrast, at low latitudes, the lack of a bow-shock structure in the western flank, possibly with a slower speed deviating inward from the modelled bow shock, may be due to the coronal background's high density and slow solar wind speed. The average Alfv{\'e}n Mach number of the shock nose is around 1.1.

(4) Coronal parameters: We derived the ambient coronal parameters with the SDR method. At distances from 7 to 13 $\rm R_{\odot}$, the Alfv{\'e}n speed, $V_A$, decreases from $\sim$890 km s$^{-1}$ to $\sim$520 km s$^{-1}$, as measured with the mean ROC of the CME nose. The coronal magnetic field strength, $B$, derived with the density model in \citet{Leblanc1998} reduces from $\sim$43 mG to $\sim$12 mG at latitudes of about -50 degrees.

In general, the expansion usually dominates in the initial phase of the CME evolution, and the rapid expansion of the CME in all directions can result in a piston-driven shock \citep{Ma2011, Cheng2012a, Ying2018}. Through a time series of 3D reconstructions of the CME front made using the graduated cylindrical shell model and the shock front using the ellipsoid model, \citet{Kwon2014} found that the 3D shock front could be a mixture of a bow shock and a piston-driven shock in the eruption 
event on 7 March 2012. In the nose direction the shock can be well represented by the ellipsoid model and behaves as a bow shock, while at the flanks the driven shock propagates in all directions and behaves as a piston-driven shock. However, the CME in this work only shows apparent motion in the nose direction and, unusually, has no noticeable lateral expansion. Based on both the evolution of the CME and the shock, there is no observational evidence supporting the existence of a piston-driven shock in the initial phase for this event.

\citet{Schmidt2016} tested the reliability of the SDR method by using 3D magnetohydrodynamics simulations for an actual regular-shaped CME and found a good agreement between the measured and simulated magnetic field strengths, with errors of 30\% from 1.8 to 10 $\rm R_{\odot}$. The result of \citet{Schmidt2016} shows support for the SDR method in a coronal environment. However, several differences between the results of our work and those of \citet{Gopalswamy2011} and \citet{Kim2012} should be pointed out: (1) In our event the averaged value of $\bar{R}_c$ is $\sim$0.5 $\rm R_{\odot}$ at the CME nose and does not change significantly during the CME propagation. In \citet{Gopalswamy2011}, $R_c$ increases continuously from 1.7 $\rm R_{\odot} $ to 3.0 $\rm R_{\odot} $ due to the CME expansion. (2) The standoff distance, $\Delta$, in our event is almost a constant $\sim$1.8 $ R_{\odot} $ (Col. 7) during the CME and shock propagation, while the $\Delta$ in \citet{Gopalswamy2011} increases gradually from 0.75 to 1.29 $ R_{\odot} $. (3) The SDR, $ \delta $, measured in this work (Col. 8) is much larger than those in \citet{Gopalswamy2011} and \citet{Kim2012}. The former has an average of $\sim$3.9, while the $\delta $ in \citet{Gopalswamy2011} and \citet{Kim2012} scatters in the range 0.19-0.78 at a height of 3 to 15 $R_{\odot}$. All these differences show the particularity of the CME in this work.

\begin{figure}[!th]
  \centering
  \includegraphics[width=0.45\textwidth]{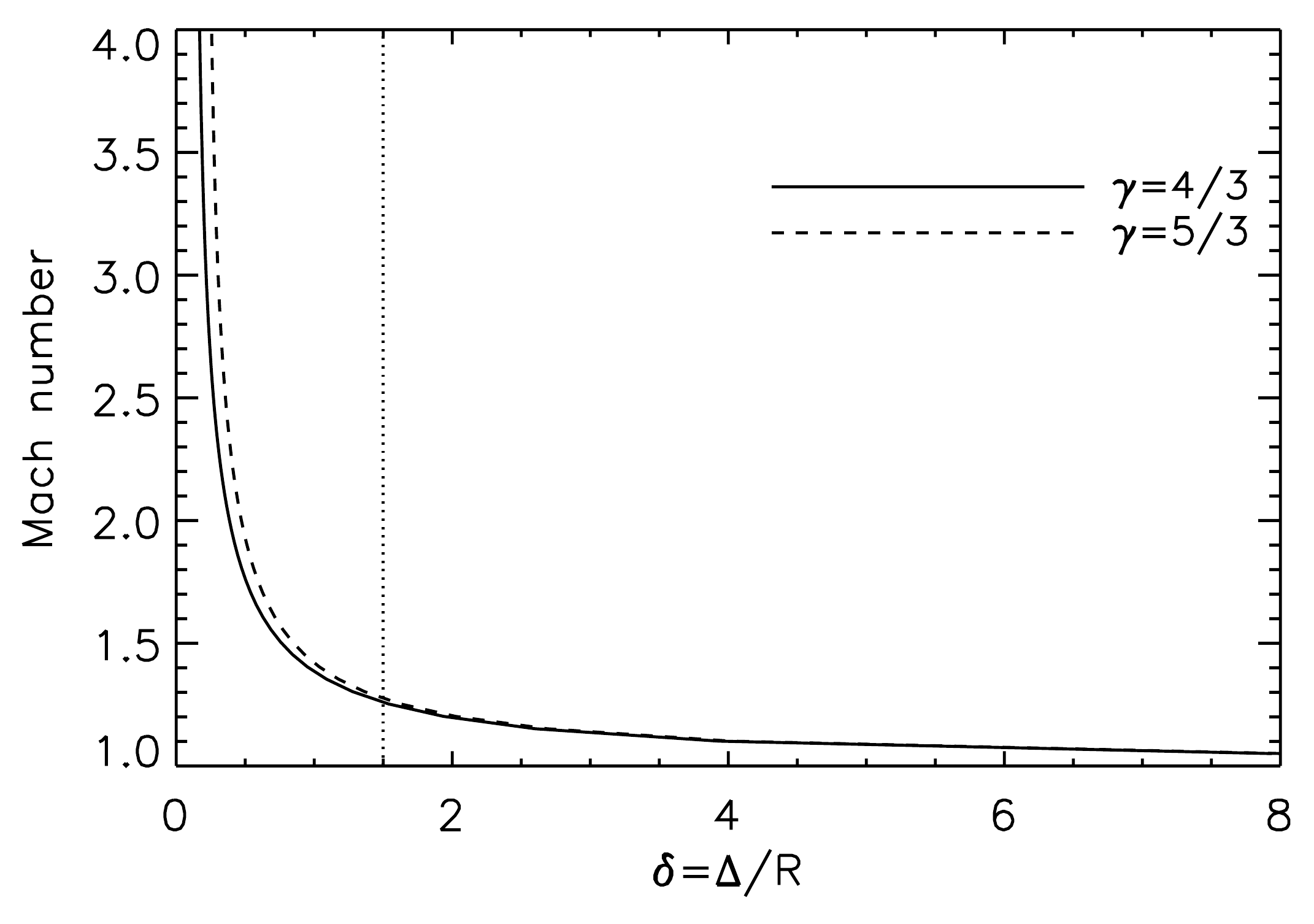}
  \caption{Alfv{\'e}n Mach number as a function of the SDR, $\delta$, according to Eq.~\ref{eq:Mach}. Solid and dashed lines denote the ratio of specific heats $\gamma =4/3$ and $5/3$, respectively. The vertical dotted line represents $\delta = 1.5$.}
  \label{fig:discuss}
\end{figure}

According to Eq.~\ref{eq:Mach}, the SDR, $\delta$, is a monotonically decreasing function of the Mach number. As $\delta$ decreases, the change range of the Mach number will vary sharply, especially when $\delta$ is less than 1.5 (such as for some blunt CMEs with large ROCs), as shown in Fig.~\ref{fig:discuss}. In addition, the role of the ratio of specific heats gradually increases when $\delta$ decreases. As we mentioned above, we find a notable discrepancy (a factor of 2-4) between the two principal ROCs of the CME nose, similar to the underestimated factor (of $3-8$) estimated by \citet{Maloney2011} due to the projection effect, which implies that simply using the CME ROC in the POS could give rise to overestimating or underestimating the coronal parameters when using the SDR method. For example, a significant difference, as considerable as that of the CME event analyzed in this article or \citet{Maloney2011}, between the minimal and maximal ROCs for a blunt aspherical CME with large ROCs and small $\delta$ ($<1.5$) can result in the extensive range of the estimated Alfv{\'e}n Mach number. Thus, multi-perspective observations and 3D reconstructions will play significant roles in the exploration of CMEs and the properties of shocks  as well as the estimation of coronal physical parameters. %

\begin{acknowledgements}
   We thank Yuandeng Shen for insightful discussions. STEREO is a project of NASA. The SECCHI data used here were produced by an international consortium of the Naval Research Laboratory (USA), Lockheed Martin Solar and Astrophysics Lab (USA), NASA Goddard Space Flight Center (USA), Rutherford Appleton Laboratory (UK), University of Birmingham (UK), Max-Planck-Institut for Solar System Research (Germany), Centre Spatiale de Li{\`e}ge (Belgium), Institut d'Optique Th{\'e}orique et Applique{\'e} (France), Institut d'Astrophysique Spatiale (France). SOHO is a mission of international cooperation between ESA and NASA. This work is supported by NSFC (grant Nos. U1731241, 11921003, 11973012), CAS Strategic Pioneer Program on Space Science (grant Nos. XDA15018300, XDA15052200, XDA15320103, and XDA15320301), the mobility program (M-0068) of the Sino-German Science Center, and the National Key R\&D Program of China (2018YFA0404200). L.F. also acknowledges the Youth Innovation Promotion Association for financial support.
\end{acknowledgements}

\bibliography{refs}
\begin{appendix}

\section{3D surface fitting of the CME bulge}

We intercepted the bulge structure of the CME, defining a Cartesian coordinate system whose Z axis passes through the $O$ point (the CME nose and the peak of the bulge) and is consistent with the symmetrical axial direction of the bow shock. X and Y define the plane perpendicular to Z. The position of the $O$ point is higher than the coordinate origin. 

\begin{figure}[h!]
    \centering
    \includegraphics[width=0.45\textwidth]{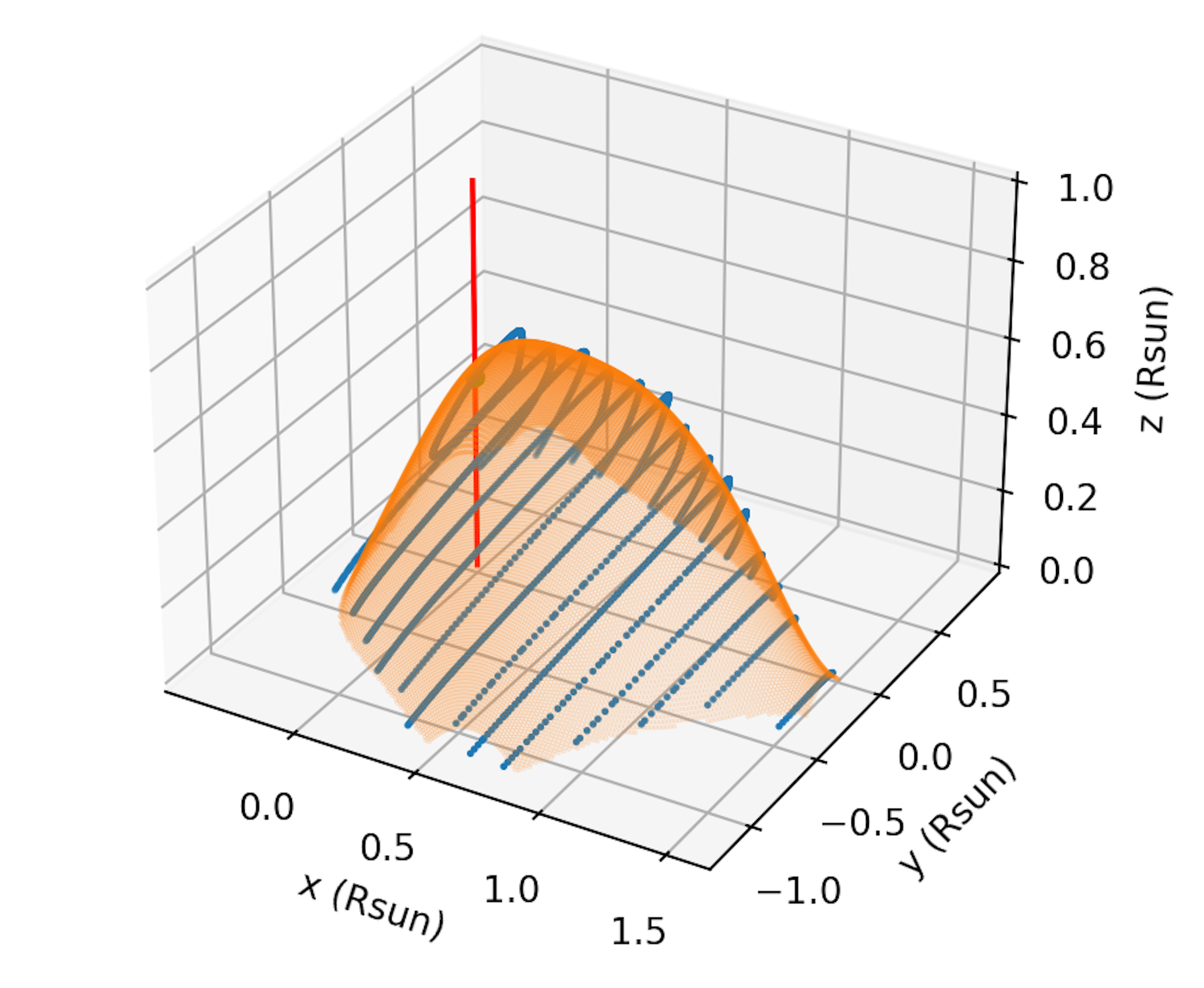}
    \caption{One example of the surface fitting of the CME bulge at 22:10 UT. Blue curves denote the cropped CME bulge reconstructed via the MF method. The orange surface is the result of the surface fitting. The red line represents the symmetrical axial direction of the bow shock. The intersection of the red line and the surface marked by a green dot is the CME nose.}
    \label{fig:append_a}
\end{figure}

We fit the surface of the CME bulge by using fifth-order polynomial functions at each time, which can be expressed as
\begin{equation}
z(x,y)=\sum_{n = 0}^{5} a[i,j,i+j\leq n]x^iy^j,
\end{equation}
where $a[i,j,i+j\leq n]$ are polynomial coefficients, which we set as $a_{ij}$, and $i$ and $j$ are non-negative integers. The fifth-order polynomial fit returns the polynomial coefficients ($a_{ij}$) and their standard deviations ($\sigma_{ij}$). Then, we used 100 Monte Carlo simulations to obtain 100 surfaces at each time. In such simulations, each polynomial coefficient, $a_{ij}$, is selected by 100 times, randomly following a Gaussian distribution with $3\sigma_{ij}$. Subsequently, we can calculate the first and second derivatives of the 100 surfaces and obtain the principal directions and ROCs at the CME nose, via Eq.~\ref{eq:h}, at each time. Monte Carlo simulations of the CME surface provide us with the uncertainty estimate. One example of the surface fitting is shown in Fig.~\ref{fig:append_a}.


\section{Calculation of radii of curvature}
\begin{figure}[htb]
\centering
\includegraphics[width=0.45\textwidth]{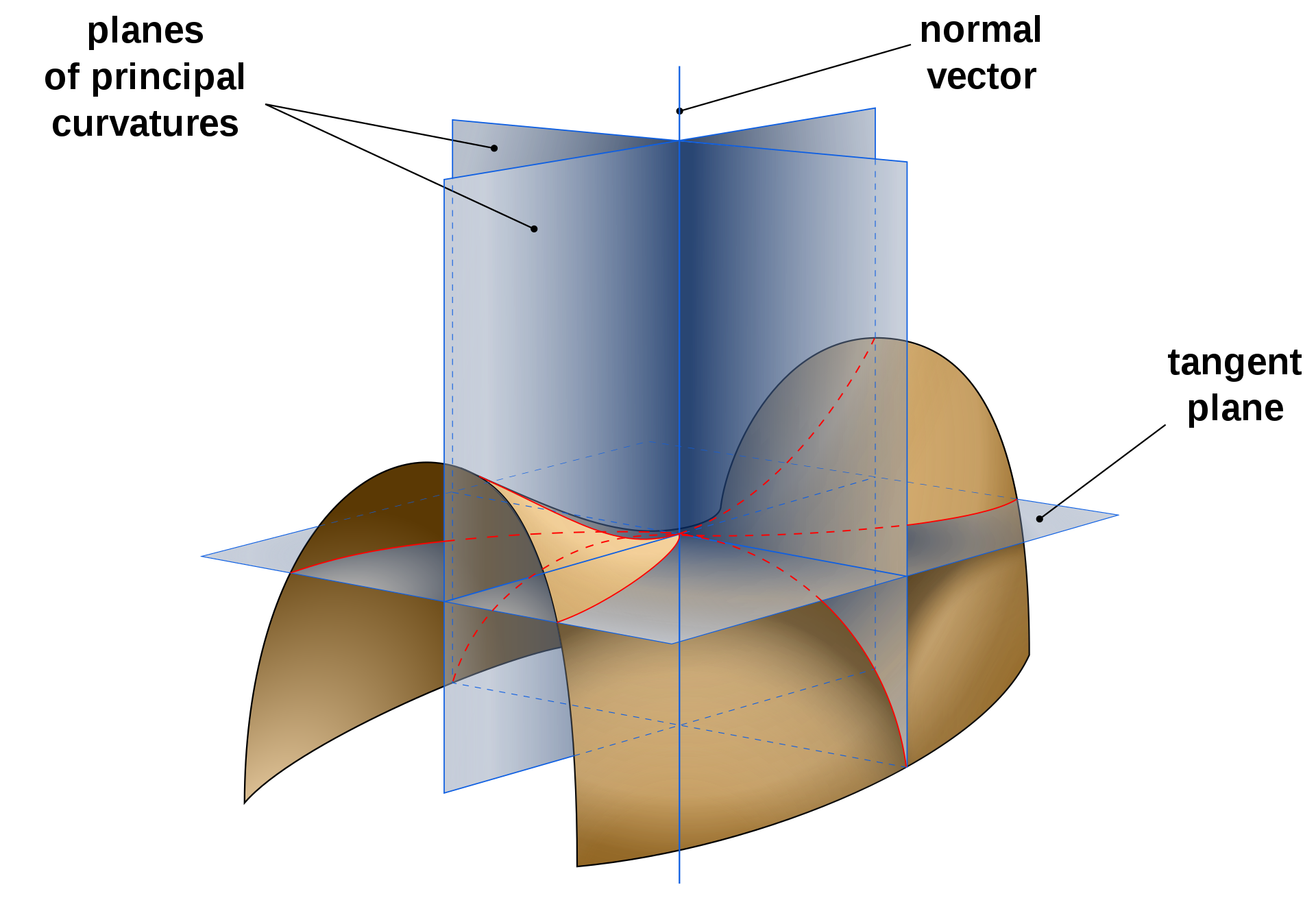}
\caption{Schematic drawing of a local surface with its tangent plane, normal vector, and two planes of principal curvatures. The two principal directions align with the intersections of the two planes of principal curvatures with the tangent plane, indicated by the blue lines on the tangent plane (https://en.wikipedia.org/wiki/Principal\_curvature\#/media/ File:Minimal\_surface\_curvature\_planes-en.svg).}
\label{fig:plane}
\end{figure}

When a local smooth surface can be expressed by the local 2D Taylor expansion of the intensity, $I$, at one cell, $\boldsymbol{i}$,
\begin{equation}
I(\boldsymbol{x})\simeq I(\boldsymbol{i})+\boldsymbol{g}^{T}(\boldsymbol{x-i})+\frac{1}{2}(\boldsymbol{x-i})^T\boldsymbol{H}(\boldsymbol{x-i}),
\end{equation}
where the elements {$\boldsymbol{g}$} and {$\boldsymbol{H}$} are
\begin{equation}
{
\left( \begin{array}{c}
g_{x} \\
g_{y} 
\end{array} 
\right )} = {
\left( \begin{array}{c}
\frac{\partial}{\partial x} \\
\frac{\partial}{\partial y}
\end{array}
\right )}I(\boldsymbol{x}),
\end{equation}

\begin{equation}
{
\left( \begin{array}{cc}
H_{xx} & H_{xy}\\
H_{yx} & H_{yy}
\end{array} 
\right )} = {
\left( \begin{array}{cc}
\frac{\partial^2}{\partial x^2} & \frac{\partial^2}{\partial x\partial y} \\
\frac{\partial^2}{\partial x\partial y} & \frac{\partial^2}{\partial y^2}
\end{array}
\right )}I(\boldsymbol{x}),
\label{eq:h}
\end{equation}
the diagonalization of $\boldsymbol{H}$ will provide us with the principal directions of the local surface and the ratio of the two principal curvatures (see Fig.~\ref{fig:plane}). The true principal curvatures need to be divided by a factor of $[1+(\frac{\partial I}{\partial x})^2+(\frac{\partial I}{\partial y})^2]^{1/2}$. They measure how much the surface bends and in which directions at that point. More details are available in Appendix A of \citet{Feng2009phd}.
\end{appendix}

%
%

\end{document}